\documentclass[12pt]{article}
\usepackage{amsmath,amssymb,epsfig}
\usepackage[vcentermath]{youngtab}

\newcommand{\comment}[1]{}

\def\swnedots{\mathinner{\mkern1mu\raise1pt\vbox{\kern7pt\hbox{.}}\mkern2mu
\raise4pt\hbox{.}\mkern2mu\raise7pt\hbox{.}\mkern1mu}}

\def\({\left(}
\def\){\right)}
\def\<{\langle}
\def\>{\rangle}

\newcommand{\Exp}{\mathrm{Exp}\,}

\newcommand{\plus}{\!+\!}

\newtheorem{prop}{Proposition}[section]
\newtheorem{theo}{Theorem}[section] 

\newenvironment{demo}[1]{\textit{Proof #1. }}{$\diamondsuit$}

\newtheorem{corol}[theo]{Corollary}
\newtheorem{defi}[prop]{Definition}
 
\numberwithin{equation}{section}

\def\surj{\to\kern-.1em\llap{$\to$}}

\def\jrus{\leftarrow\kern-.1em\llap{$\leftarrow$}}
\def\strictsubset{\hbox{$\subseteq\kern-.4em\llap{${}_/$}$}}

\begin{document}
\title{{Discrete  Polynomials and Discrete Holomorphic Approximation}}
\author{Christian \textsc{Mercat}\\
Technische Universit\"at Berlin, Germany\\
\href{mailto:mercat@sfb288.math.tu-berlin.de}{mercat@sfb288.math.tu-berlin.de}}
\maketitle
\begin{abstract} We use discrete holomorphic polynomials to prove that, given
  a refining sequence of critical maps of a Riemann surface, any holomorphic
  function can be approximated by a converging sequence of discrete
  holomorphic functions.
\end{abstract}

\section{Introduction}
The notion of discrete Riemann surfaces has been defined in~\cite{M01}.  We
are interested in discrete surfaces given by a cellular decomposition
$\diamondsuit$ of dimension two, where all faces are \emph{quadrilaterals}.
We suppose $\diamondsuit$ bipartite and it defines (away from the boundary)
two dual cellular decompositions $\Gamma$ and $\Gamma^*$, edges $\Gamma^*_1$
are dual to edges $\Gamma_1$, faces $\Gamma^*_2$ are dual to vertices
$\Gamma_0$ and vice-versa.  Their union is denoted the \emph{double}
$\Lambda=\Gamma\sqcup \Gamma^*$. A \emph{discrete conformal structure} on
$\Lambda$ is a real positive function $\rho$ on the unoriented edges
satisfying $\rho(e^*)=1/\rho(e)$. It defines a genuine Riemann surface
structure on the discrete surface: Choose a length $\delta$ and realize each
quadrilateral by a lozenge whose diagonals have a length ratio given by
$\rho$.  Gluing them together provides a flat riemannian metric with conic
singularities at the vertices, hence a conformal structure~\cite{Tro}.  This
data leads to a straightforward discrete version of the \emph{Cauchy-Riemann
  equation}.  A function on the vertices of $\diamondsuit$ is discrete
holomorphic iff for every quadrilateral $(x,y,x',y')\in\diamondsuit_2$,
\begin{equation}
  \label{eq:CR}
  f(y')-f(y)=i\, \rho(x,x')\left( f(x')-f(x)\right).
\end{equation}
\begin{figure}[htbp]
\begin{center}\input{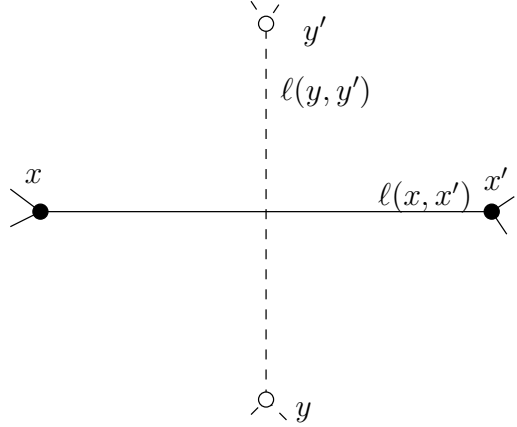}
\end{center}
\caption{The discrete Cauchy-Riemann equation.}         \label{fig:CR}
\end{figure}

Given a fixed flat riemannian metric on a Riemann surface, with a finite
number of conic singularities, we define a discrete conformal structure as
\emph{critical} if the flat riemannian metric it gives rise to is isometric
to this fixed one. Let's stress that this notion is useful mostly when
considering a sequence of discrete conformal structures, all adapted to the
same fixed flat metric. A \emph{refining} sequence is such a sequence of
critical maps where the common side lengths $\delta_k$ of the diamonds in the
map $\diamondsuit_k$, form a sequence converging to zero, and where the lozenge
angles are bounded away from zero (the faces don't collapse).

In~\cite{M01} we proved that 1) any Riemann surface accepts a refining
sequence of critical maps and 2) a converging sequence of discrete
holomorphic functions on a refining sequence of critical maps converges to a
continuous holomorphic function. It is the main purpose of this article to
prove the converse result:
\begin{theo}\label{th:approx}
  Given a refining sequence of critical maps of a compact flat simply
  connected surface $U$, any holomorphic function on $U$ can be approximated
  by a converging sequence of discrete holomorphic functions.
\end{theo}
The proof is based on series expansion. In Sec.~\ref{sec:Integration} we
define discrete integration at criticality and show that the sequence of
primitives of converging discrete holomorphic functions converge to the
continuous primitive. We define discrete polynomials and use them to prove
the main theorem.

In Sec.~\ref{sec:Derivation}, we define discrete derivation and present its
differences with the continuous one. It allows none the less for the
coefficients of a discrete polynomial to be computed by successive
derivation. In the annex, we discuss different aspects of the discrete theory
which differ or don't appear in the continuous case, in
Sec.~\ref{sec:FaceDeriv} another notion of derivation, in
Sec.~\ref{sec:Basis} the analog of the Leibnitz formulae, in
Sec.~\ref{sec:Minimal} a motivation of the derivation formula through a study
of the minimal polynomial and in Sec.~\ref{sec:Series} a discussion of the
problems arising when considering discrete series, with the example of the
discrete exponential.

\section{Integration at criticality} \label{sec:Integration}
Away from a conic singularity, a critical map locally forms a flat partition
of the plane by lozenges. The relevance of this kind of maps in the context
of discrete holomorphy was first pointed out and put to use by
Duffin~\cite{Duf68}. He defined the discrete analogues of the integer powers
of $Z$ and the derivation formula that we are going to give.

The crucial point about criticality is the following remark: Given an
isometric local map $Z:U\cap\Lambda\to\mathbb{C}$, where the image of the
quadrilaterals are lozenges in $\mathbb{C}$, any holomorphic function
$f\in\Omega(\Lambda)$ gives rise to an holomorphic $1$-form $f dZ$ defined by
the formula,
\begin{equation}
  \label{eq:fdZ}
  \int_{(x,y)} f dZ :=\frac{f(x)+f(y)}2 \(Z(y)-Z(x)\),
\end{equation}
where $(x,y)\in\diamondsuit_1$ is an edge of a lozenge. It is averaged into a
holomorphic $1$-form on $\Lambda$:
\begin{equation}
  \label{eq:fdZLambda}
  \int_{(x,x')} f dZ=\frac{f(x)+f(y)+f(x')+f(y')}4 \(Z(x')-Z(x)\).
\end{equation}

Alternatively, it provides a way to integrate a function by taking the
primitive of $f dZ$.
\begin{theo}\label{th:primitive}
  Given a sequence of discrete holomorphic functions $(f_k)$ on a refining
  sequence of critical maps, converging to a holomorphic function $f$, the
  sequence of primitives $(\int f_k\, dZ)$ converges to $\int f(z)\, dz$.
  Moreover, in the compact case, if the convergence of the functions is of
  order $O(\delta_k^2)$, it stays this way for the primitives.
\end{theo}

\begin{demo}{\ref{th:primitive}}
  We recall from~\cite{M01} that we extend a discrete holomorphic function
  $f_d$ from the vertices $\diamondsuit_0$ linearly to the edges
  $\diamondsuit_1$ and harmonicly to the faces $\diamondsuit_2$ to obtain a
  continuous piecewise harmonic function $\widehat f_d$ of the surface. We
  proved that, given a refining sequence $(\diamondsuit^k)$ of critical maps,
  the point-wise convergence of a sequence of discrete holomorphic functions
  $(f_k)$, restricted to the sequence of vertices $(\diamondsuit^k_0)$,
  implies a uniform limit of the continuous extensions $(\widehat f_k)$ to a
  genuine (continuous) holomorphic function $f$.
  
  Let $U$ be the flat simply connected patch under consideration.  We suppose
  that we are given a sequence of flat vertices $O_k\in\diamondsuit_k$ where
  the face containing the fixed flat origin $O\in U$ is adjacent to $O_k$.
  Let $\widehat F_k$ the extension of the primitive $\int_{O_k} f_k\, dZ$ to
  the whole surface. We want to prove that the following sequence tends to
  zero
\begin{equation}
  \label{eq:uniformF}
\(\left|(\widehat F_k(x)-\widehat F_k(O))-\int_O^x f(z)\, 
dz\right|\)_{k\in\mathbb{N}}.
\end{equation}

Let us fix a point $x$, and for each integer $k$ consider a vertex
$x_k\in\diamondsuit_0$ on the boundary of the face of $\diamondsuit_2$
containing $x$.

We decompose the difference~\eqref{eq:uniformF} into three parts, inside the
face containing the origin $O$ and its neighbor $O_k$, similarly for $x$ and
$x_k$, and purely along the edges of the graph $\diamondsuit^k$ itself.
\begin{eqnarray}
|\bigl(\widehat F_k(x)&-&\widehat F_k(O)\bigr)-\int_O^x f(z)\, dz| \;=\;\notag\\&&
\left|(\widehat F_k(x)-\widehat F_k(x_k))+\int_{O_k}^{x_k}f_k\, dZ+
(\widehat F_k(O)-\widehat F_k(O_k))-
\int_O^x f(z)\, dz\right|\notag\\
&\leq&
\left|\widehat F_k(x)-\widehat F_k(x_k)-\int_{x_k}^x f(z)\, dz\right|+
\left|\int_{O_k}^{x_k}f_k\, dZ-\int_{O_k}^{x_k} f(z)\, dz\right|+\notag\\
&&\left|\widehat F_k(O_k)-\widehat F_k(O)-\int_{O}^{O_k} f(z)\, dz\right|.  
\label{eq:F}
\end{eqnarray}
On the face of $\diamondsuit$ containing $x$, the primitive $x\mapsto
\int_{x_k}^x f(z)\, dz$ is a holomorphic, hence harmonic function as well as
$x\mapsto \widehat F_k(x)$.  By the maximum principle, the harmonic function
$x\mapsto \widehat F_k(x)-\widehat F_k(x_k)-\int_{x_k}^x f(z)\, dz$ reaches
its maximum on that face, along its boundary. The difference of the
discrete primitive along the edges of $\diamondsuit$ is equal to
\begin{equation}
\widehat F_k((1-\lambda)x+\lambda y)-\widehat
F_k(x)=\lambda(y-x)\frac{f_k(x)+f_k(y)}2
\label{eq:Fdifint}
\end{equation}
while because $f$ is differentiable with a bounded derivative on $U$,
\begin{eqnarray}
  \int_{x}^{(1-\lambda)x+\lambda y} f(z)\, dz&=&
\lambda(y-x)\frac{f(x)+f(y)}2+(y-x)^2\frac{\lambda^2f'(x)+(1-\lambda)^2f'(y)}4
+o(\delta_k^3)\notag\\
&=&\lambda(y-x)\frac{f(x)+f(y)}2+O(\delta_k^2)
  \label{eq:fint}
\end{eqnarray}
so that
\begin{equation}
  \label{eq:Fx}
  |\widehat F_k(x)-\widehat F_k(x_k)-\int_{x_k}^x f(z)\, dz|=O(\delta_k^2).
\end{equation}
  Similarly for the term around the origin.
  
  By definition of $\widehat f_k$, the $1$-form $\widehat f_k(z)\, dz$ along
  edges of the graph $\diamondsuit$ is equal to the discrete form $f_k dZ$ so
  that $\int_{O_k}^{x_k}f_k\, dZ=\int_{O_k}^{x_k}\widehat f_k(z)\, dz$ on a
  path along $\diamondsuit$ edges.
  Therefore the difference
\begin{equation}
  \label{eq:Fint}
\left|\int_{O_k}^{x_k}f_k\, dZ-\int_{O_k}^{x_k} f(z)\, dz\right|\leq
\int_{O_k}^{x_k}\left|\(\widehat f_k(z)-f(z)\)\, dz\right|
\end{equation}
is of the same order as the difference $\left|f_k(z)-f(z)\right|$ times the
length $\ell(\gamma_k)$ of a path on $\diamondsuit_k$ from $O_k$ to $x_k$.
We showed in~\cite{M01} that provided that the lozenge angles are in the
interval $(\eta,2\pi-\eta)$ with $\eta>0$ (the faces don't collapse), this
length can be bounded as $\ell(\gamma_k)\leq \frac 4{\sin\eta}|x_k-O_k|$.
Since we are interested in the compact case, this length is bounded uniformly
and the difference~\eqref{eq:Fint} is of the same order as the point-wise
difference.  We conclude that the sequence of discrete primitives converges to
the continuous primitive and if the limit for the functions was of order
$O(\delta^2)$, it remains of that order.
\end{demo}

Following Duffin~\cite{Duf, Duf68}, we define by inductive integration the
discrete analogues of the integer power monomials $z^k$, that we denote
$Z^{:k:}$:
\begin{eqnarray}
  \label{eq:Zk}
  Z^{:0:}&:=&1,\\
  Z^{:k:}&:=&\int_{O} \,Z^{:k-1:}\, dZ,
\end{eqnarray}
where $O$ is a fixed flat origin. The choice of this origin is relevant and
we present in the Appendix the formulae for the change of basis of discrete
polynomials from one base point to another. Although polynomials are
translated into polynomials of same degree, the formulae are slightly more
involved than the Leibnitz rule $(Z-a)^n=\sum_{k=0}^n \binom{k}{n}Z^k
a^{n-k}$. When sequences of maps are involved, we suppose that we have chosen
a flat origin for each map, forming a converging sequence to a flat point.
For example we suppose that a fixed flat point $O$ on the surface is a vertex
of each map.

The discrete polynomials of degree less than three agree point-wise with their
continuous counterpart, $Z^{:2:}(x)=Z(x)^2$.

A simple induction then gives the following
\begin{corol}\label{cor:poly}
  The discrete polynomials converge to the continuous ones, the limit is
  of order $O(\delta_k^2)$.
\end{corol}

Which implies the main theorem:
\begin{demo}{\ref{th:approx}}
  On the simply connected compact set $U$, a holomorphic function $f$ can be
  written, in a local map $z$ as a series,
  \begin{equation}
    \label{eq:series}
    f(z)=\sum_{k\in\mathbb{N}} a_k z^k.
  \end{equation}
  Therefore, by a diagonal procedure, there exists an increasing integer
  sequence $\(N(n)\)_{n\in\mathbb{N}}$ such that the sequence of discrete
  holomorphic polynomials converge to the continuous series. 
\begin{equation}
  \label{eq:seriesPoly}
  \(\sum_{k=0}^{N(n)} a_k Z^{:k:}\)_{n\in\mathbb{N}}\to f.
\end{equation}
\end{demo}

\section{Derivation at criticality} \label{sec:Derivation}
We are interested in getting the coefficients of a polynomial by the same
process as in the continuous case: successive derivation. We are going to see
that this operation, while possible, is more difficult than in the
continuous.

Let $\varepsilon$ be the biconstant $\varepsilon(\Gamma)=+1$,
$\varepsilon(\Gamma^*)=-1$. The main issue is to separate this ''mode'' from
the constant mode in order to define properly the discrete analog of the
value of a function at the origin.

On a finite map, the dimension of the space of discrete holomorphic functions
is equal to half of the number of boundary points plus one,
\begin{equation}
  \label{eq:boundaryBase}
  \Omega(U)\sim \mathbb{C}^{|\partial U|/2+1},
\end{equation}
for example the values on the boundary points $\Gamma^*_0\cap\partial U$ on
the graph $\Gamma^*$ and in one point $O\in\Gamma$.  The dimension of the
space of discrete polynomials can not be greater than this number. This
implies the existence of a minimal polynomial
\begin{equation}
P_Z=\sum_{k=1}^na_k\,Z^{:k:}\equiv 0.\label{eq:PZ}
\end{equation}
 For a holomorphic function $f$, the equality
$f\,dZ\equiv 0$ is equivalent to $f=\lambda\,\varepsilon$ for some
$\lambda\in \mathbb{C}$, we can uniquely normalize $P_Z$ so that
$\sum_{k=1}^nk\,a_k\,Z^{:k-1:}=\varepsilon$, that is $a_1=1$.

We can therefore take as a basis of the discrete polynomials the set 
\begin{equation}
  \label{eq:basisPoly}
  \(1,Z,Z^{:2:},\ldots,Z^{:n-2:},\varepsilon\).
\end{equation}

We conjectured in~\cite{M01} that when $U$ is convex~\cite{M0111043},
\eqref{eq:basisPoly} is a basis of the whole space of discrete holomorphic
functions. In any case, one can supplement them by an orthogonal complement
and we will denote $\nu\in\mathbb{C}^{|\partial U|/2+1}$ the vector encoding
the linear form yielding the coordinate along the constant function $1$ in
this basis:
\begin{eqnarray}
  \nu\cdot\sum_{k=0}^{n-2}\,a_k\,Z^{:k:}&=&a_0,\notag\\
  \nu\cdot\varepsilon&=&0, 
 \label{eq:nu}\\
\nu\cdot f&=&0\;\; \text{ ~when~ }\;\; \forall k,\;\; Z^{:k:}\cdot f=0.
\notag
\end{eqnarray}
Notice that since the change of base point for polynomials is triangular in
the basis of monomials, the vector $\nu$ does not depend on the map $Z$ or on
the base point but only on the discrete conformal structure. An explicit
value for this vector would be desirable, especially for numerical purposes.

Following Duffin~\cite{Duf, Duf68} (see Sec.\ref{sec:Minimal} for a
motivation), we introduce the
\begin{defi}
  For a holomorphic function $f$, define on a flat simply connected map $U$
  the holomorphic functions $f^\dag$, the \emph{dual} of $f$, and $f'$, the
  \emph{derivative} of $f$, by the following formulae:
  \begin{equation}
f^\dag(z):=\varepsilon(z)\,\bar f(z),\label{eq:fdag}
\end{equation}
where $\bar f$ denotes the
  complex conjugate, $\varepsilon=\pm 1$ is the biconstant, and
  \begin{equation}
f'(z):=\frac4{\delta^2}\left( \int_{O}^z f^\dag
    dZ\right)^\dag+\lambda\,\varepsilon,\label{eq:deffp}
  \end{equation}
with $\lambda$ determined by
\begin{equation}
  \label{eq:fix}
  \lambda=\nu \cdot f'.
\end{equation}
\end{defi}
If, in the context of sequences, the function $f$ can be Taylor expanded
around the origin, $f(x)=f(O)+\mu\, x+O(x^2)$, then $\lambda=\mu$ and the
identity~\ref{eq:fix} is equivalent in the $O(x^2)$ sense to 
\begin{equation}
  \label{eq:fixApproxHarmoFace}
  \sum_k \rho(O,x_k)\(\frac{f(y_{k+1})-f(y_k)}{Z(y_{k+1})-Z(y_k)}-\lambda\)=0,
\end{equation}
where $(O,y_k,x_k,y_{k+1})\in\diamondsuit_2$ are the quadrilaterals adjacent
to the origin. Therefore this condition can be used for numerical purposes in
replacement of the exact derivation formula~\ref{eq:fix}. It states that the
derivative at the origin is the mean value of the nearby face derivatives
(see Sec.~\ref{sec:FaceDeriv}). For example, in the rectangular lattice
$\diamondsuit=\delta\,(\mathbb{Z}e^{i\,\theta}+\mathbb{Z}e^{-i\,\theta})$, with
horizontal parameter $\rho=\tan\theta$ and vertical parameter its inverse, it
provides the good choice for the first three and all the even degrees,
$\lambda_0=0$, $\lambda_1=1$ and $\lambda_{2k}=0$, but yields $\lambda_k= k!
\(\frac{\delta}{2}\)^{k-1}\cos (k-1)\theta=O(\delta^{k-1})$ for $k$ odd and
fixed, instead of $0$ for $k>2$.


\begin{prop}\label{prop:fp}
  The derivative $f'$ fulfills
\begin{equation}
  \label{eq:dffpdz}
  d_\diamondsuit f = f' dZ.
\end{equation}
  The discrete monomials verify
\begin{equation}
  \label{eq:monomial}
 \forall\, k<n-1,\qquad (Z^{:k:})'=k\,Z^{:k-1:}.
\end{equation}
\end{prop}
Eq.\eqref{eq:dffpdz} was proved in~\cite{M01} and Eq.\eqref{eq:monomial}
follows from
\begin{equation} \label{eq:epsdZ}
  f\,dZ\equiv 0\;\Leftrightarrow f=\lambda\,\varepsilon.\;\diamondsuit
\end{equation}

\appendix

\section{Face derivation} \label{sec:FaceDeriv}
Another derivation operator can be defined. By definition of holomorphy, for
each face of the graph $\diamondsuit$, the $1$-form $d_\Lambda f$ is
proportional to $dZ$ along its pair of dual diagonals. This defines a linear
operator
\begin{equation}
 \frac{d}{dZ}:\Omega(\Lambda\cap U)\to
C^2(\diamondsuit\cap U).\label{eq:fpC2}
\end{equation}
Given the local flat map $Z$, the space
$C^2(\diamondsuit\cap U)$ can be seen as the space $C^{(1,0)}(\Lambda\cap U)$
of type $(1,0)$ forms on $\Lambda$, the $+i$-eigenvectors of the Hodge star
$*$, that is to say the forms $\alpha$ such that, for dual edges
$(y,y')=(x,x')^*$,
\begin{equation}
  \label{eq:10}
  \int_{(y,y')}\alpha=i\rho(x,x') \int_{(x,x')}\alpha.
\end{equation}
It defines 
\begin{equation}
a_Z\in C^2(\diamondsuit\cap U)\;\text{~ for ~}\;\alpha=a_Z dZ
\label{eq:funcD2}
\end{equation}
We can say that the face function $a_Z$ is holomorphic whenever the
corresponding form $\alpha$ is closed, that is to say by imposing the Morera
theorem. This idea was developed by Kenyon~\cite{Ken02}. Notice that the
$1$-form $\alpha$ is then co-closed as well. If the quadrilateral graph
$\diamondsuit$, along with its discrete conformal structure was isomorphic to
its dual graph $\diamondsuit^*$, a choice of isomorphy could define a
derivation endomorphism $\frac{d\;}{dZ}:\Omega(\Lambda\cap
U)\to\Omega(\Lambda\cap U)$.  But for this to happen, the vertices have to
have degree four, so although important, the only case is $\mathbb{Z}^2$ with
the choice of a translation by $(n+\frac12,m+\frac12)$, $n,m\in\mathbb{Z}$.

It is immediate that given a $O(\delta^2)$-converging sequence of discrete
holomorphic functions, this face derivation yields a sequence of functions,
extended on each face as piece-wise constant functions, which converges to
the continuous derivative.

\section{Change of basis}\label{sec:Basis}

We are now going to consider the change of coordinate.  Let $Z$ a critical
map.  If $\zeta$ is another critical map with $\zeta=a\,(Z-b)$ on their
common definition set, the change of map for $Z^{:k:}$ is not as simple as
the Leibnitz rule $\left(a\,(z-b)\right)^{k}=a^{k}\sum_{j=0}^{k}\binom{k}{j}
z^{k-j}(-b)^{j}$ but is a deformation of it.  The problem is that pointwise
product is not respected, $Z^{:k+\ell:}(z)\neq Z^{:k:}(z)\times
Z^{:\ell:}(z)$. In particular the first is holomorphic and not the second.
For each partition of $m=k+\ell$ into a sum of integers there is a
corresponding monomial of degree $m$.  An inference shows that the result is
still a polynomial in $Z$:
\begin{prop}\label{prop:Young}
  The powers $\zeta^{:k:}$ of the translated critical map $\zeta=a(Z-b)$ are
  given by
    \begin{equation}
        \zeta^{:k:}=a^{k}\sum_{j=0}^{k}\binom{k}{j}(-1)^{j}Z^{:k-j:}B^{j}(b)
        \label{eq:yngSum}
    \end{equation}
    where  $B^{j}(b)$ corresponding to 
    $b^{j}$ is a  sum over all the degree $j$ monomials in 
    $b$, defined recursively by $B^{0}=1$ and
    \begin{equation}
    B^{k}(b):=\sum_{j=0}^{k-1}\binom{k}{j}(-1)^{k+j+1}Z^{:k-j:}(b)B^{j}(b)
        \label{eq:yngBk}
    \end{equation}
\end{prop}
\begin{demo}{\ref{prop:Young}}
  The calculation is easier to read using Young diagrams. Denote the
  point-wise product of monomials
\begin{equation}
    \left(Z^{:k_{1}:}(z)\right)^{\ell_{1}}
    \left(Z^{:k_{2}:}(z)\right)^{\ell_{2}} \ldots
    \left(Z^{:k_{n}:}(z)\right)^{\ell_{n}}
    \label{eq:yngZkl}
\end{equation}
with $k_{1}>k_{2}>\ldots >k_{n}$ as a Young diagram $Y$, coding columnwise
the partition of the integer given by the total degree
$k=\sum_{j=1}^{n}k_{j}\times\ell_{j}$ into the sum of
$\ell=\sum_{j=1}^{n}\ell_{j}$ integers.  For example the following monomial
of degree $15=3\times 2+2\times 4+1$ is noted \Yboxdim{.5em}
\begin{equation}
    \left(Z^{:3:}(z)\right)^{2}
    \left(Z^{:2:}(z)\right)^{4}  
    Z^{:1:}(z)=:\;\yng(7,6,2).
    \label{eq:yng}
\end{equation}
Then, $B^{j}(b)=\sum_{Y}c(Y)Y(b)$ where the sum is over all Young diagrams of
total degree $j$, $c(Y)$ is an integer coefficient that we are going to
define and $Y(b)$ is the pointwise product of the monomials coded by $Y$ at
$b$.  The coefficient of the Young diagram $Y$ above is given by the
multinomials
\begin{equation}
   c(Y)=(-1)^{k+\ell}\frac{k!}{(k_{1}!)^{\ell_{1}}(k_{2}!)^{\ell_{2}}\cdots 
   (k_{n}!)^{\ell_{n}}}\;\frac{\ell!}{\ell_{1}!\ell_{2}!\cdots \ell_{n}!}.
   \label{eq:yngCoeff}
\end{equation}
\end{demo}

For example, the first Young diagrams have the coefficients \Yboxdim{.5em}
$c({\scriptsize\young(\hfil\hfil n\hfil\hfil)})=\frac{ n!}{1!\cdots
  1!}\frac{n!}{n!}=n!$, $c({\scriptsize
  \young(\hfil,\hfil,n,\hfil,\hfil)})=(-1)^{n+1}$,
$c({\scriptsize\young(\hfil\hfil,\hfil,n,\hfil,\hfil)})=
(-1)^{n+1}\frac{(n+1)!}{n!\,1!}\,\frac{2!}{1!\, 1!}=(-1)^{n+1}2{(n+1)}$ and
$c({\scriptsize \young(\hfil\hfil
  n\hfil\hfil,\hfil)})=-\frac{(n+1)!}{2!}\,\frac{n!}{1!\,(n-1)!}=
-\frac{(n+1)!\;n}{2}$.  The first few terms are listed explicitly in
Table~\ref{tbl:Bk}.

It is to be noted that the formula doesn't involve the shape of the graph,
the integer coefficients for each partition are universal constants and add
up to $1$ in each degree.  As a consequence, since in the context of a
refining sequence of critical maps,
\begin{equation}
Z^{:k:}(z)Z^{:\ell:}(z)=Z^{:k+\ell:}(z)+O(\delta^{2})\label{eq:ZkZlOdelta2}
\end{equation}
with $k, \ell, z$ fixed, the usual Leibnitz rule is recovered in
$O(\delta^{2})$.  Let's stress again that these functions $B^{k}$ are
discrete functions on the graph $\Lambda$ which \emph{are not discrete
  holomorphic}.
\begin{table}[tbp]\Yboxdim{.5em}
    \centering
    \begin{eqnarray}
        B^{0} & = & +\;1
        \notag  \\
        B^{1} & = & +\;\yng(1)
        \nonumber  \\
        B^{2} & = & -\;\yng(1,1)+2\;\yng(2)
        \nonumber  \\
        B^{3} & = & +\;\yng(1,1,1)-6\;\yng(2,1)+6\;\yng(3)
        \nonumber  \\
        B^{4} & = & 
        -\;\yng(1,1,1,1)+8\;\yng(2,1,1)+6\;\yng(2,2)-36\;\yng(3,1)+24\;\yng(4)
        \nonumber  \\
        B^{5} & = & 
        +\;\yng(1,1,1,1,1)-10\;\yng(2,1,1,1)-20\;\yng(2,2,1)+60\;\yng(3,1,1)+
        90\;\yng(3,2)-240\;\yng(4,1)+120\;\yng(5)
        \nonumber  \\
        B^{6} & = & 
        -\;\yng(1,1,1,1,1,1)+12\;\yng(2,1,1,1,1)+30\;\yng(2,2,1,1)
        -90\;\yng(3,1,1,1)+20\;\yng(2,2,2)-360\;\yng(3,2,1)+480\;\yng(4,1,1)
        \nonumber\\
        &&-90\;\yng(3,3)+1080\;\yng(4,2)-1800\;\yng(5,1)+720\;\yng(6)
        \nonumber
    \end{eqnarray}
    \caption{The first analogs of $z^{k}$ needed in a change of basis.}
    \label{tbl:Bk}
\end{table}

\section{Minimal polynomial}\label{sec:Minimal}
We give a motivation to Duffin's derivation formula Eq.~\eqref{eq:deffp} by
studying the minimal polynomial, its behavior regarding duality $\dagger$ and
its derivatives. It shows that derivation in $Z$ acts as integration in
$Z^\dagger$.

Consider the minimal polynomial
\begin{equation}
P_Z=\sum_{k=1}^na_k\,Z^{:k:}\equiv 0.\label{eq:PU}
\end{equation}
with $a_1=1$, or equivalently
$\sum_{k=1}^nk\,a_k\,Z^{:k-1:}=\varepsilon=1^\dag$. This implies that
\begin{eqnarray}
\sum_{k=2}^n k(k-1)\,a_k\,Z^{:k-2:}
&=&\frac4{\delta^2}\,Z^\dag+a_2\,2\,\varepsilon,
\label{eq:varepsderiv}\\
\sum_{k=3}^n\frac{k!}{(k-3)!}\,a_k\,Z^{:k-3:}
&=&\(\frac4{\delta^2}\)^2\frac{{Z^{:2:}}^\dag}2
+\,a_2\,\frac4{\delta^2}\,2\,Z^\dag+\,a_3\,3!\,1^\dag,
\notag\\
\sum_{k=\ell}^n\frac{k!}{(k-\ell)!}\,a_k\,Z^{:k-\ell:}
&=&\sum_{k=0}^{\ell-1}a_{\ell-k}\(\frac4{\delta^2}\)^{k}
\frac{(\ell-k)!}{k!}\,{Z^{:k:}}^\dag,
\notag\\
0&=&\sum_{k=0}^{n}a_{n+1-k}\(\frac4{\delta^2}\)^{k}
\frac{(n+1-k)!}{k!}\, {Z^{:k:}}^\dag.
\notag\end{eqnarray}

Since $P_Z(Z)^\dag=\bar P_Z(Z^\dag)$, a symmetry among the coefficients
follows:
\begin{equation}
  \label{eq:akanp1mk}
  a_{k}=\frac{(n+1-k)!}{n!k!}\;
\frac{\bar a_{n+1-k}}{\bar a_n}\(\frac4{\delta^2}\)^{k-1},
\end{equation}
leading to
\begin{equation}
  \label{eq:akanp1mk2}
  k!\,|a_{k}|\(\frac4{\delta^2}\)^{\frac{n-2k+1}4}=
(n+1-k)!|a_{n+1-k}|\(\frac4{\delta^2}\)^{\frac{n-2(n+1-k)+1}4},
\end{equation}
in particular $|a_n|=\frac1{(n!)}\,\(\frac4{\delta^2}\)^{\frac{n-1}2}$ and when
$n$ is odd, $\text{arg~}(a_{\frac{n+1}2})=\frac12\text{arg~}(a_n)$.

\section{Series} \label{sec:Series}
We present here some problems linked to discrete series. They can converge
only when their coefficients decrease at least exponentially fast. And
exponentially fast is enough since one can define a discrete exponential.


The $O(\delta^2)$ convergence of discrete polynomials to continuous ones is
in fact slower and slower as the degree of the polynomial grows.  Consider
for example $\diamondsuit$ containing the chain
$\{0,\frac{1}{n},\frac{2}{n},\ldots ,1\}$, the first $Z^{:k:}(x)$ with
$x=\frac{\ell}{n}$ are listed in Table~\ref{tbl:Zkx}. And in general, a
scaling argument shows that 
\begin{equation}
     |Z^{:k:}(x)-x^{k}|\leq \lambda_{k}|x|^{k-2}\delta^{2}.
     \label{eq:Zkxk}
 \end{equation}
\begin{table}[tbp]
    \centering
    \begin{equation*}
        \!\!\!\!\begin{array}{c|c|c|c|c|c|}
            k & 3 & 4 & 5 & 6 & 7  \\
            \hline&&&&&\\
            Z^{k}(x) & x^{3}\plus\frac{x}{2n^{2}} & x^{4}\plus\frac{2x^{2}}{n^{2}} & 
            x^{5}\plus\frac{5x^{3}}{n^{2}}\plus\frac{3x}{2n^{4}}& 
            x^{6}\plus\frac{10x^{4}}{n^{2}}\plus\frac{23x^{2}}{n^{4}} & 
            x^{7}\plus\frac{35x^{5}}{n^{2}}\plus\frac{49x^{3}}{n^{4}}\plus
            \frac{45x}{4n^{6}}  \end{array}
    \end{equation*}
    \caption{The first powers on the interval $[0,1]/n$ for $x=\ell/n$.}
    \label{tbl:Zkx}
\end{table}
But the constant $\lambda_k$ is growing exponentially fast with $k$.  It is
for example not true that, at a point close enough to the origin, $Z^{:k:}$
will tend to zero with increasing $k$. On the contrary, if
$x=\delta\,e^{i\,\theta}$ is a neighbor of the origin with
$(O,x)\in\diamondsuit_{1}$ and $k\geq 1$, then
\begin{equation}
Z^{:k:}(x)=\frac{k!}{2^{k-1}} \,x^{k}\label{eq:Zkneighbor}
\end{equation}
in fact diverges with
$k$.  If $y$ is a next neighbor of the origin, with the rhombi
$(O,x,y,x')\in\diamondsuit_{2}$ having a half angle $\theta$ at the origin,
\begin{equation}
Z^{:k:}(y)=\frac{k!}{2^{2k-2}} \frac{\sin k\, \theta}
{\sin\theta \cos^{k-1}\theta} \,y^{k}\label{eq:Zknext}
\end{equation}
has the same diverging behavior, and so has every point at a given finite
distance of the origin.  It is only in the scaling limit with the proper
balance given by criticality that one recovers the usual behavior $
|x|<1\implies |x^{k}|\xrightarrow[k\to\infty]{} 0$.


Therefore, for its general term to converge to zero, the coefficients of a
discrete series must decrease at least exponentially fast.  The discrete
exponential $\Exp({:}\lambda{:}\,Z)$, for $|\lambda|\not=2/\delta$, is
defined by
\begin{eqnarray}
\Exp({:}\lambda{:}\,O)&=&1\label{eq:Exp}\\
d\,\Exp({:}\lambda{:}\,Z)&=&\lambda\,\Exp({:}\lambda{:}\,Z)\,dZ\label{eq:dExp}
\end{eqnarray}
This discrete holomorphic function was first defined in~\cite{M0111043} and
put to a very interesting use in~\cite{Ken02}.  The discrete exponential is a
rational fraction in $\lambda$ at every point,
\begin{equation}
  \label{eq:ExpRatFrac}
  \Exp({:}\lambda{:}\,x)=\prod_{k}
\frac{1+\frac{\lambda\delta}{2}e^{i\,\theta_k}}{
1-\frac{\lambda\delta}{2}e^{i\,\theta_k}}
\end{equation}
where $(\theta_k)$ are the angles defining $(\delta\,e^{i\,\theta_k})$, the
set of ($Z$-images of) $\diamondsuit$-edges between
$x=\sum\delta\,e^{i\,\theta_k}$ and the origin. Expanding
$\exp(\log(\Exp({:}\lambda{:}\,x)))$ in $O(\delta^2)$, for
$|\lambda|<2/\delta$ and $x$ fixed in a refining sequence of critical maps,
we get that
\begin{equation}
\Exp({:}\lambda{:}\,x)=\exp(\lambda\, x)+O(\delta^2).\label{eq:ExpOdelta2}
\end{equation}
For $|\lambda|<2/\delta$, the series
$\sum_{k=0}^{\infty}\frac{\lambda^{k}\,Z^{:k:}}{k!}$ is equal to
\eqref{eq:ExpRatFrac} whenever the former is defined, as its exterior
derivative fulfills the right equation~\eqref{eq:dExp} and its value at the
origin is $1$.  The great difference with the continuous case is that the
series is absolutely convergent only for bounded parameters,
$|\lambda|<\frac{2}{\delta}$. This suggests that asking for a product such
that
\begin{equation}
\Exp({:}\lambda{:}\,Z)\cdot\Exp({:}\mu{:}\,Z)=\Exp({:}\lambda+\mu{:}\,Z),\label{eq:prod}
\end{equation}
or equivalently $Z^{:k:}\cdot Z^{:\ell:}= Z^{:k+\ell:}$ may not be the right
choice. The symmetry $\dag$ is interesting as well,
\begin{equation}
  \label{eq:Expdag}
  \Exp({:}\lambda{:}\,Z)^\dag=\Exp({:}\frac{1}{\bar\lambda}{:}\,Z),
\end{equation}
in particular, $ \Exp({:}\infty{:}\,Z)=\varepsilon$.

The general change of basis of a given series however possible in theory is
nevertheless complicated and the information on the convergence of the new
series is difficult to obtain. The exponential remains a particularly simple
case, if $\zeta=a(Z-b)$:
\begin{equation}
    \sum_{k=0}^{\infty}\frac{\lambda^{k}}{k!}\zeta^{:k:}\propto
    \sum_{k=0}^{\infty}\frac{(a\,\lambda)^{k}}{k!}Z^{:k:}.
    \label{eq:ExpMapChgSeries}
\end{equation}

\section*{Acknowledgements}                     \label{sec:Acknowledgements}
I thank Trevor Welsh for simplifying Eq.~\eqref{eq:yngCoeff}.  This research
is supported by the Sonderforschungsbereich 288.

\bibliographystyle{unsrt}
 \bibliography{Poly}

\end{document}